\documentclass[10pt,preprint]{emulateapj}

\newcommand{\cbo}{CB$190$}
\newcommand{\spitzer}{\textit{Spitzer}}
\newcommand \msun{\hbox{$\hbox{M}_{\odot}$}}

\begin{document}

\title{\textit{Spitzer} observations of a 24$\micron$ shadow: Bok
Globule CB190 \altaffilmark{1}}

\author{Amelia M. Stutz\altaffilmark{2}, John
  H. Bieging\altaffilmark{2}, George H. Rieke\altaffilmark{2}, Yancy
  L. Shirley\altaffilmark{2}, Zoltan Balog\altaffilmark{2}, Karl
  D. Gordon\altaffilmark{2}, Elizabeth M. Green\altaffilmark{2},
  Jocelyn Keene\altaffilmark{3}, Brandon C. Kelly\altaffilmark{2},
  Mark Rubin\altaffilmark{3}, Michael W. Werner\altaffilmark{3}}

\altaffiltext{1}{This work is based in part on observations made with
the \textit{Spitzer Space Telescope}, which is operated by the Jet
Propulsion Laboratory, California Institute of Technology, under NASA
contract 1407.}  

\altaffiltext{2}{Department of Astronomy and Steward Observatory,
University of Arizona, 933 North Cherry Avenue, Tucson, Arizona 85721;
astutz@as.arizona.edu.}

\altaffiltext{3}{Jet Propulsion Lab, California Institute of
Technology, 4800 Oak Grove Drive, Pasadena, CA 91109.}

\begin{abstract}

We present {\it Spitzer} observations of the dark globule CB190
(L771).  We observe a roughly circular 24$\micron$ shadow with a
$70\arcsec$ radius.  The extinction profile of this shadow matches the
profile derived from 2MASS photometry at the outer edges of the
globule and reaches a maximum of $\sim\!32$ visual magnitudes at the
center.  The corresponding mass of CB190 is $\sim\!10$~$\msun$.  Our
$^{12}$CO and $^{13}$CO J = 2-1 data over a
10$\arcmin\times$10$\arcmin$ region centered on the shadow show a
temperature $\sim\!10$~K.  The thermal continuum indicates a similar
temperature for the dust.  The molecular data also show evidence of
freezeout onto dust grains.  We estimate a distance to CB190 of 400~pc
using the spectroscopic parallax of a star associated with the
globule.  Bonnor-Ebert fits to the density profile, in conjunction
with this distance, yield $\xi_{max} = 7.2$, indicating that \cbo\ may
be unstable.  The high temperature (56~K) of the best fit Bonnor-Ebert
model is in contradiction with the CO and thermal continuum data,
leading to the conclusion that the thermal pressure is not enough to
prevent free-fall collapse.  We also find that the turbulence in the
cloud is inadequate to support it.  However, the cloud may be
supported by the magnetic field, if this field is at the average level
for dark globules.  Since the magnetic field will eventually leak out
through ambipolar diffusion, it is likely that \cbo\ is collapsing or
in a late pre-collapse stage.

\end{abstract}


\section{Introduction}

Cold cloud cores, where star formation begins, represent the stage in
early stellar evolution after the formation of molecular clouds and
before the formation of Class 0 objects.  Their emission is
inaccessible at shorter wavelengths, such as the near infrared and
visual bands, due to low temperatures, very high gas densities, and
associated large amounts of dust.  Because cold cloud cores can best
be observed at sub-millimeter and far-infrared wavelengths, these
spectral regions are essential to developing an understanding of the
first steps toward star-formation \citep[see,
e.g.,][]{bacmann00,kirk07}.  The wavelength range accessible to the
{\it Spitzer} Space Telescope, 3.6~$\micron$ to 160~$\micron$, is
ideally suited to observe cold, dense regions.

\cbo\ (L771) is an example of one such dark globule and is classified
in the Lynds catalog as having an opacity of 6 \citep{lynds62},
i.e., very high.  \citet{clemens88} study this object as part of an
optically selected survey of small molecular clouds.  They find that
it appears optically isolated, is somewhat asymmetric ($a/b\sim2.5$)
and has some bright rims of reflection and H$\alpha$.  \cbo\ is
$\sim\!5\arcmin$ across, and has an estimated distance of 400~pc
\citep{neckel80}.

We present Spitzer maps of \cbo.  In particular, we highlight the
observation of this globule in absorption at 24~$\micron$.  We combine
these data with SCUBA observations at 850~$\micron$.  We have obtained
complementary Heinrich Hertz Telescope (HHT) $^{12}$CO and $^{13}$CO
J=2-1 on the fly (OTF) maps of this globule and have used the HHT and
Green Bank Telescope (GBT) to measure high resolution line profiles
for $^{12}$CO, $^{13}$CO, NH$_3$, CCS, C$_3$S, and HC$_5$N.  We report
C$^{18}$O and DCO$^+$ measurements with the Caltech Submillimeter
Observatory (CSO).  We discuss the issue of stability and possible
support mechanisms in some detail; however, we cannot say conclusively
if \cbo\ is in equilibrium.  In \S~2 we describe the observations and
data processing.  In \S~3 we present our main analysis of \cbo: we
derive an optical depth and an extinction profile for the 24~$\micron$
shadow using a technique presented for the first time in this
work\footnote{Previous related work has been conducted using the ISO
7$\micron$ band, \citep[e.g.,][]{bacmann00}}.  We also discuss two
sources associated with \cbo\ and derive a distance estimate.  In \S~4
we compare various mass estimates of this object.  In \S~5 we describe
our Bonnor-Ebert fitting method and discuss possible support
mechanisms for \cbo.  Finally, in \S~6 we summarize our main
conclusions.  All positions are given in the J2000 system.

\section{Observations and processing}

\subsection{Spitzer data}

Object \cbo\ (L$771$), centered at about RA = $19^h 20^m 48^s$, Dec =
$+23^o 29\arcmin 45\arcsec$, was observed with the MIPS instrument
\citep{rieke04} at $24$~$\micron$, $70$~$\micron$ and $160$~$\micron$,
\spitzer\ program ID $53$ (P.I. G. Rieke).  The observations were
carried out in scan map mode.  Figure~\ref{fig:mips} shows these data,
along with the Digital Sky Survey (DSS) red plate image of the
globule.

The 24~$\micron$ data were reduced using version 3.06 of the MIPS Data
Analysis Tool (DAT; Gordon et al.\ 2005).  In addition to the standard
processing, several other additional steps have been applied: 1)
correcting for variable offsets among the four readouts; 2) applying
a scan-mirror position-dependent flat field; 3) applying a scan-mirror
position-independent flat field to remove long term gain changes due
to previous saturating sources; and 4) background subtraction.  For
the last two steps, masks of any bright sources and also of the region
of interest were used to ensure that the two corrections were
unbiased.  The background subtraction was performed by fitting a low
order polynomial to each scan leg of the masked data and subtracting
the resulting fit.  This procedure removes the contribution of
zodiacal and other background light as well as small transients seen
after a boost frame.

The 70~$\micron$ and 160~$\micron$ data were also reduced using the
DAT (Gordon et al. 2005).  After completing the standard reduction
processing, an additional correction was performed to remove the long
term drift in the Ge:Ga detectors.  The correction was determined by
fitting a low order polynomial to the masked version of the entire
dataset for each pixel.  By masking bright sources as well as the
region of interest we ensure that the masked-version fits are
unbiased.  The resulting fits for each pixel were subtracted to remove
the long term drift as well as any background light.

\begin{figure}
  \epsscale{0.85}\plotone{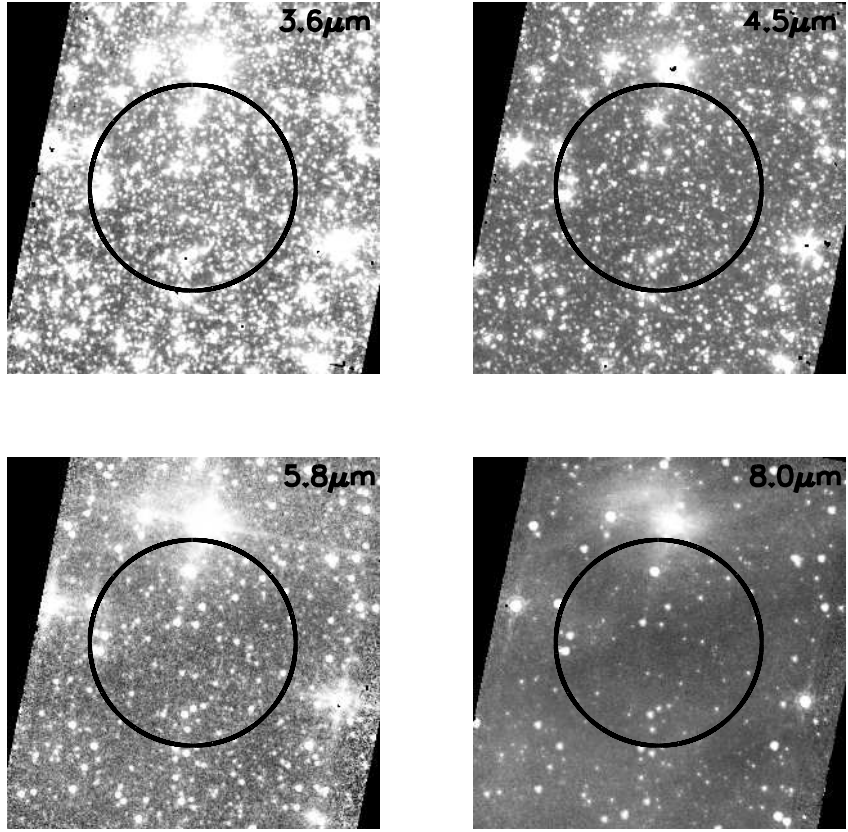}
  \caption{Gallery of IRAC \cbo\ images centered on the $24\micron$
    shadow: RA = $19^h 20^m 48^s$, Dec = $+23^o 29\arcmin 45\arcsec$.
    The images are oriented such that north is up and east is to the
    left, they are $6\arcmin$ on a side, and the corresponding
    effective wavelengths of the four IRAC channels are labeled.  The
    original mosaiced IRAC image pixel scale is $0\farcs6$ pix$^{-1}$
    and the images displayed here are shown at $1\farcs2$ pix$^{-1}$.
    The black circles are centered on the 24$\micron$ shadow and have
    a radius of 100$\arcsec$.}
  \label{fig:irac}
\end{figure}

\cbo\ was observed with the IRAC instrument \citep{fazio04} at $3.6$,
$4.5$, $5.8$ and $8.0$~$\micron$, program ID $94$ \
(P.I. C. Lawrence), see fig.~\ref{fig:irac}.  Standard packages were
used to reduce the data, and the mosaicked frames were generated with
the MOPEX software package.  The data were taken in high dynamic range
mode; the $30$ second exposure time was divided into a $1.0$s
``short'' frame and a $26.8$~s ``long'' frame at each position.  Each
observation was repeated $5$ times, yielding an effective long-frame
exposure time of $134$~s.  SExtractor \citep{bertin96} was used for
both source extraction and photometry.  The photometry was
cross-checked with PhotVis version $1.1$, an IDL GUI-based
implementation of DAOPHOT \citep{gutermuth04}.  We found good
agreement between the two sets of photometry.

\subsection{$^{12}$CO and $^{13}$CO data}

The \cbo\ region was mapped in the J=2-1 transitions of $^{12}$CO and
$^{13}$CO with the 10-m diameter HHT on Mt. Graham, Arizona on 2005
June 9.  The receiver was a dual polarization SIS mixer system
operating in double-sideband mode with a 4 - 6 GHz IF band.  The
$^{12}$CO J=2-1 line at 230.538 GHz was placed in the upper sideband
and the $^{13}$CO J=2-1 line at 220.399 GHz in the lower sideband,
with a small offset in frequency to ensure that the two lines were
adequately separated in the IF band.  The spectrometers, one for each
of the two polarizations, were filter banks with 1024 channels of 1
MHz width and separation.  At the observing frequencies, the spectral
resolution was 1.3~km~s$^{-1}$ and the angular resolution of the
telescope was 32$\arcsec$ (FWHM).

A $10\arcmin \times 10\arcmin$ field centered at RA = $19^h 20^m
49.5^s$, Dec = $+23^o 29\arcmin 57\arcsec$ was mapped with on-the-fly
(OTF) scanning in RA at $10\arcsec$~sec$^{-1}$, with row spacing of
$10\arcsec$ in declination, over a total of 60 rows.  This field was
observed twice, each time requiring about 100 minutes of elapsed time.
System temperatures were calibrated by the standard ambient
temperature load method \citep{kutner81} after every other row of the
map grid.  Atmospheric conditions were clear and stable, and the
system temperatures were nearly constant at $T_{sys} = 450$~K (SSB).

Data for each polarization and CO isotopomer were processed with the
{\it CLASS} reduction package (from the University of Grenoble
Astrophysics Group), by removing a linear baseline and convolving the
data to a square grid with $16\arcsec$ grid spacing (equal to one-half
the telescope beamwidth).  The intensity scales for the two
polarizations were determined from observations of DR21(OH) made just
before the OTF maps.  The gridded spectral data cubes were processed
with the {\it Miriad} software package \citep{sault95} for further
analysis.  The two polarizations were averaged, yielding images with
rms noise per pixel and per velocity channel of 0.15~K-T$_A ^*$ for
both the $^{12}$CO and $^{13}$CO transitions.

\begin{figure}
  \begin{center}
    \scalebox{0.45}{{\includegraphics{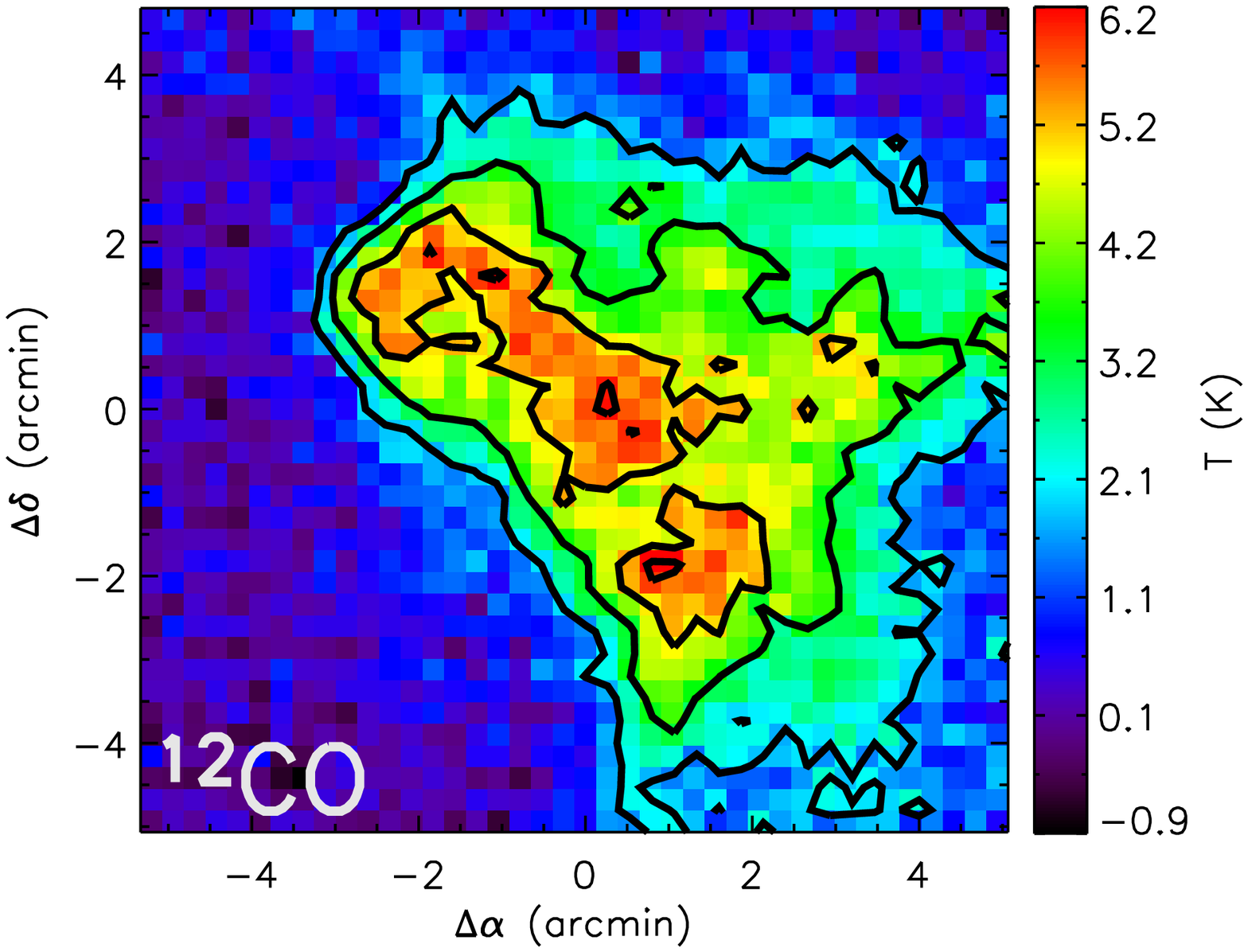}}{\includegraphics{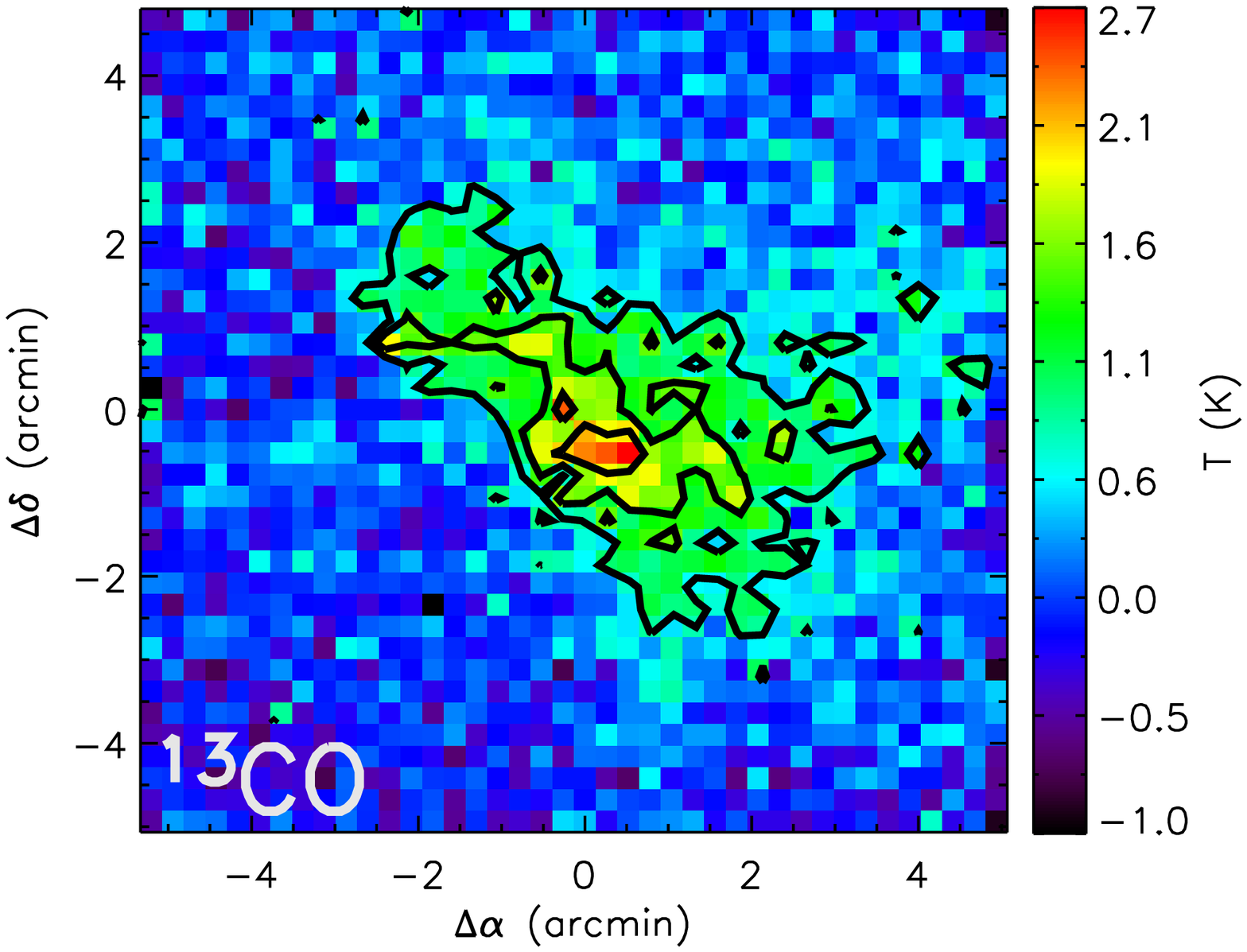}}}
    \caption{Maps of the integrated $^{12}$CO (J = 2-1, $\nu$ =
    220.399 GHz) and $^{13}$CO (J = 2-1, $\nu$ = 230.538 GHz) lines
    taken with a FWHM telescope resolution of $32\arcsec$ and
    convolved to a square grid cell spacing of $16\arcsec$.  The
    central coordinates of the images are RA = $19^h 20^m 49^s$, Dec =
    $+23^o 29\arcmin 57\arcsec $.  The color-bars indicate the
    temperature scales in each map.  The $^{12}$CO contour levels are
    2, 3.5, 5, and 6 K, and the $^{13}$CO contour levels are 0.8, 1.4,
    and 2.0 K.  }
    \label{fig:co}
  \end{center}
\end{figure}

The linewidths were narrow, so that 70\% of the flux in the $^{12}$CO
J=2-1 line was in a single 1 MHz spectrometer channel, while
essentially all the flux of the $^{13}$CO line was in a single
channel.  We can therefore set an upper limit of
$\sim\!1.3$~km~s$^{-1}$ on the linewidth for the emission lines, but
have little or no kinematic information from the maps, other than the
LSR velocity, which is 11.0 km~s$^{-1}$, in agreement with
\citet{clemens88}.  In figure~\ref{fig:co} we show two maps of the
integrated $^{12}$CO and $^{13}$CO J=2-1 lines, summed over the 2
spectrometer channels with detectable emission.  Furthermore, in
figure~\ref{fig:overbw}, we show the $^{13}$CO J=2-1 contours
overlayed on the 24~$\micron$ image; the spatial coincidence between
the two is evident.  

\begin{figure}
  \epsscale{0.65}\plotone{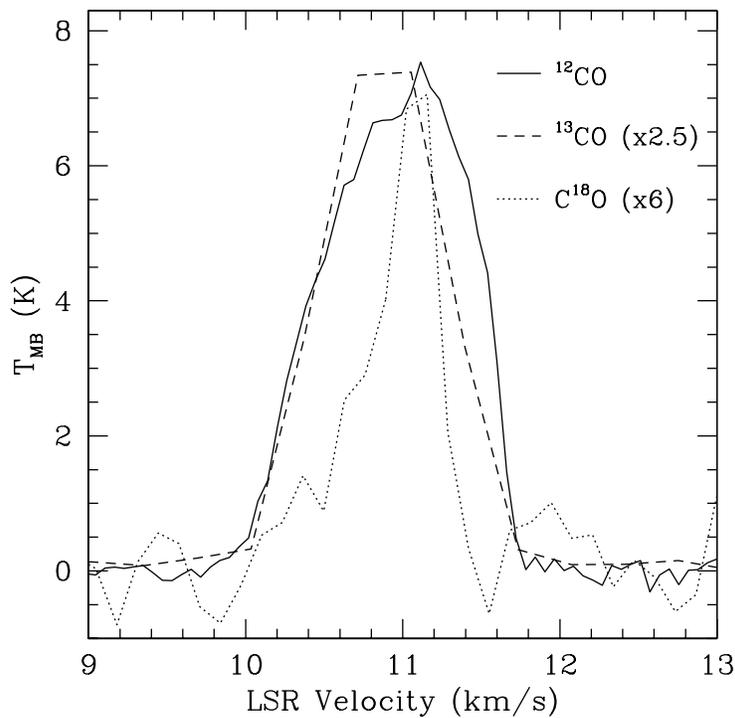}
  \caption{High resolution spectra of $^{12}$CO and $^{13}$CO J = 2 -
    1 lines at the position of the peak CO optical depth, which
    coincides with the center of the 24~$\mu$m shadow.  The spectral
    resolution of the $^{13}$CO line is 0.25 MHz or 0.34~km~s$^{-1}$,
    and the resolution of the $^{12}$CO line is 0.05~MHz or
    0.061~km~s$^{-1}$ (R = 5,000,000).  The vertical scale of the
    $^{13}$CO spectrum is multiplied by 2.5.  Total integration time
    was 15~min on source.  RMS noise in the $^{12}$CO spectrum is
    0.12~K (main-beam brightness temperature.  The dotted line shows
    the C$^{18}$O J = 2 - 1 spectrum obtained with the Caltech
    Submillimeter Observatory at a spectral resolution of 0.1~MHz or
    0.13~km~s$^{-1}$ (Michael M. Dunham - private communication,
    2006).  The vertical scale of the C$^{18}$O spectrum is multiplied
    by 6.}
  \label{fig:hr12}
\end{figure}

To better constrain the CO line properties, we observed the core of
the molecular cloud with high velocity resolution on 2006 June 21 with
the HHT.  The position observed was at RA = $19^h 20^m 46.4^s $, Dec =
$+23^o 29\arcmin 45\farcs6 $, which is the peak of the $^{12}$CO
intensity map (fig.~\ref{fig:co}).  Both the $^{12}$CO and $^{13}$CO
J=2-1 transitions were observed for 15 minutes each.  The high
resolution $^{12}$CO and $^{13}$CO spectra are shown in
fig.~\ref{fig:hr12}.  The $^{12}$CO line is slightly asymmetric, with
a peak beam-averaged brightness temperature of 7.65~K, and a velocity
width (FWHM) of 1.22~km~s$^{-1}$.  The $^{13}$CO line is narrower,
with a FWHM of 0.97~km~s$^{-1}$, and has a peak intensity of 2.96~K.
The $^{12}$CO/$^{13}$CO intensity ratio at the line peak is therefore
$\sim\!2.6$; if the isotopic ratio [$^{12}$CO/$^{13}$CO] = 50, the
optical depth of the $^{13}$CO line at the peak is $\sim\!0.5$, and
the line is optically thin.  While values for [$^{12}$CO/$^{13}$CO] in
the range of 50 to 70 are reasonable, see e.g., \citet{milam05},
changing the isotopic ratio will not significantly affect our
calculated optical depth.  We can then compute an integrated CO column
density assuming the CO rotational levels are in LTE.  Following
\citet{rohlfs04}, the peak $^{12}$CO line brightness temperature
implies a CO excitation temperature of 12.6~K.  Assuming this applies
to both isotopomers, the integrated $^{13}$CO J=2-1 line intensity
gives an integrated column density of N($^{13}$CO)$ = 2 \times
10^{15}$~cm$^{-2}$ at the cloud peak, and N($^{12}$CO) $ = 9 \times
10^{16}$~cm$^{-2}$.  If the [CO/H$_2$] abundance ratio were $1.5
\times 10^{-4}$, typical of molecular clouds, the column density of
H$_2$ would be N(H$_2$) $ = 6 \times 10^{20}$~cm$^{-2}$.  A standard
gas to dust ratio and extinction law would then imply $A_V$ = 0.7~mag
through the cloud core.  This very low implied extinction is clearly
incompatible with the observed large extinction evident in the POSS
image from which the L771 dark cloud was identified.  This discrepancy
suggests that the CO molecule is substantially depleted by freeze-out
onto dust grains in the core of the cloud, as has been seen in many
other molecular cloud cores (e.g., Tafalla et al. 2002).  We confirm
this conclusion in \S~4.3.

\subsection{Observations of other molecular lines}

Observations of CB190 were performed with the 105 meter Green Bank
Telescope on September 20, 2006.  Five spectral lines were observed
simultaneously in dual polarization: NH$_3$ (1,1) and (2,2), CCS $N_J
= 1_2 - 2_1$, C$_3$S $J = 4 - 3$, and HC$_5$N $J = 9 - 8$.  The
correlator was set up with 6.1~kHz resolution and 8 spectral windows
($4 RR$ and $4 LL$ polarizations) with a 50~MHz bandpass.  The peak
$^{13}$CO position was observed for 20 minutes of ON-source
integration time while frequency switching with a frequency throw of
4.15~MHz.  The atmospheric optical depth at 1.3~cm was calibrated
using the weather model of Ron Maddalena (private communication,
2006).  The average opacity was $\tau_{1.3} = 0.095 \pm 0.010$ during
the CB190 observations.  The main beam efficiency was determined from
observations of the quasars 3C286 and 3C48 and was $\eta_{mb} = 0.75
\pm 0.04$.  The observations were reduced using standard GBTIDL script
for frequency-switched observations, calibrated using the latest
K-band receiver T$_{cal}$ for each polarization, and corrected for
atmospheric opacity and the main beam efficiency.

Michael M. Dunham (private communication, 2006) provided C$^{18}$O J =
2 - 1 and DCO$^+$ J = 3 - 2 observations obtained with the Caltech
Submillimeter Observatory.  The observations were performed in
position-switching mode with the 50 MHz AOS backend.  The main beam
efficiency was measured to be 0.74 during the observing run.
C$^{18}$O J = 2 - 1 was detected toward the 24~$\mu$m shadow peak
position, but DCO$^+$ was not detected to a 3 $\sigma$ rms of 0.4~ K.

The molecular line emission detected toward CB190 is striking in its
lack of diversity.  Only CO isotopomers and NH$_3$ have been detected
to date.  The early-time molecules CCS, C$_3$S, and HC$_5$N were not
detected with the GBT to a 25~mK (T$_R^*$) baseline rms while the
deuteration tracer, DCO$^+$ was not detected in the CSO observations.
The NH$_3$ (1,1) line is weak, with a peak T$_R^* = 200 \pm 23$~mK and
a narrow linewidth of $\Delta v = 0.47 \pm 0.08$~km~s$^{-1}$.  Since
the main line is a blend of many hyperfine lines, the actual linewidth
is smaller.  The optical depth is low enough that the satellite lines
are barely detected at the $3 \sigma$ level.  Assuming an excitation
temperature of 10~K and optically thin emission, the column density of
NH$_3$ is $N = 5.4^{+0.6}_{-0.4} \times 10^{12}$~cm$^{-2}$.  This
column density is two orders of magnitude below the median column
density in the NH$_3$ survey of \citet{jijina99} and is six times
lower than their weakest detection.  A modest 20 minute integration
time on the GBT can probe very low column densities of NH$_3$.
Unfortunately, the (2,2) line was also not detected to the 23~mK
baseline rms level; therefore, we cannot independently determine the
kinetic temperature of the gas.  The extremely weak NH$_3$ emission
and DCO$^+$ non-detection may indicate that CB190 is a relatively
young core.  In contrast, there is evidence that CO is depleted
indicating that the core is not a nascent dense core. A more extensive
and sensitive search for molecular line emission should be attempted
toward CB190 to characterize its chemistry.

\subsection{2MASS Data}

We used the 2MASS All-Sky Point Source Catalog (PSC)
\citep{skrutskie06} photometry.  We quote the default J, H, and K-band
photometry in this work, labeled j$_{-}$m, h$_{-}$m, and k$_{-}$m in
the 2MASS table header.  The magnitudes are derived over a 4$\arcsec$
radius aperture.  We use the combined, or total, photometric
uncertainties for the default magnitudes, labeled j$_{-}$msigcom,
h$_{-}$msigcom, and k$_{-}$msigcom in the 2MASS table header.

\subsection{SCUBA Data}

\begin{figure}
  \epsscale{0.65}\plotone{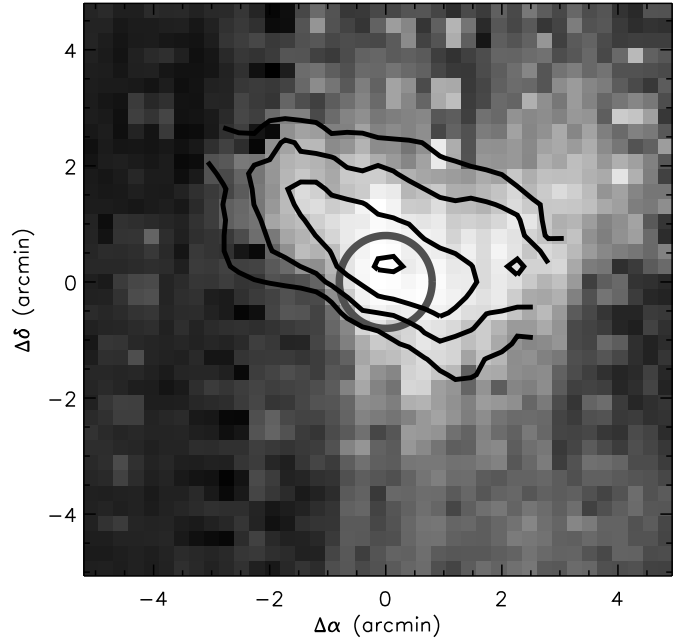}
  \caption{160$\micron$ image detail, centered on the 24$\micron$
      shadow, with 850$\micron$ contours overlayed in black.  The
      160$\micron$ image is shown at the original mosaic pixel scale
      of $16\farcs0$~pix$^{-1}$.  The 850$\micron$ contour levels are
      0.1, 0.2, 0.3 and 0.4~Jy~beam$^{-1}$, and the beam is a
      $32\arcsec$ FWHM Gaussian.  The grey circle indicates the
      location of the 24 $\micron$ shadow and is 48$\arcsec$ in
      radius.}
  \label{fig:overscubw}
\end{figure}

\begin{figure}
  \epsscale{0.95}\plotone{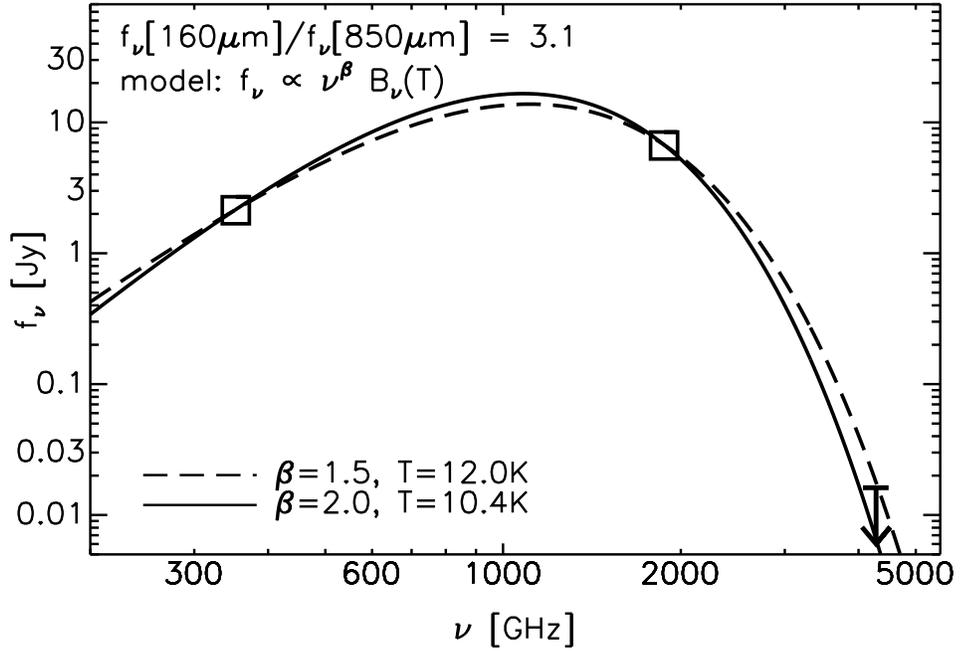}
  \caption{70$\micron$, 160$\micron$, and 850$\micron$ photometry
    using a 48$\arcsec$ radius aperture centered on the 24$\micron$
    shadow coordinates.  We show two models for the dust emission with
    the indicated assumed values of $\beta$ and the corresponding
    best-fit temperatures.  }
  \label{fig:scuphot}
\end{figure}

We include in this work the \citet{visser02} reduced 850$\micron$
SCUBA map of \cbo\ (Claire Chandler, private communication, 2006).
This map was convolved with a 32$\arcsec$ FWHM Gaussian beam and is
shown in figure~\ref{fig:overscubw} as contours overlayed on the
160~$\micron$ data.  The spatial agreement between the two wavelengths
is very good.  We note, however, that the southern edge of the cloud
may be artificially sharpened at 850$\micron$ due to the position
angle of chopping during the scan map.  Figure~\ref{fig:scuphot} shows
photometry for the 70~$\micron$, 160~$\micron$ and 850~$\micron$
observations, measured with a 48$\arcsec$ radius aperture centered on
the 24~$\micron$ shadow coordinates.  The observed ratio of the long
wavelength fluxes is $f_\nu[160\micron]/f_\nu[850\micron] = 3.1$.
With this ratio we fit for a cloud temperature using the model
\begin{equation}
f_\nu \propto \nu^\beta B_\nu(T).
\end{equation}
We fix the value of $\beta$ at 1.5 and 2.0, a reasonable range of
values for dust emissivity \citep{whittet92}, and derive best-fit
model temperatures of 12.0~K and 10.4~K, respectively.  These two
models are plotted in figure~\ref{fig:scuphot}.  The 160~$\micron$
flux is well detected, with a signal to noise ratio $\sim\!20$,
indicating that these temperatures are robust.  However, to test the
models conservatively, we recalculate model temperatures allowing for
20\% errors in our photometry.  We do not find significant variation in
the derived temperatures.  We note that the lack of a detection at
70~$\micron$ mildly favors the colder temperature of 10.4 K.  The $3
\sigma$ upper limit to the flux density at 70~$\micron$ is
$\sim\!16$~mJy while the predicted values are 17~mJy for the $\beta =
1.5$, T = 12~K model, and 6~mJy for the $\beta = 2.0$, T = 10.4~K
model.

\citet{visser02} find that the \cbo\ SCUBA 450~$\micron$ emission is
spatially offset from the 850~$\micron$ data for reasons that are not
understood.  Because we find good spatial agreement between the
850~$\micron$ data and the 160~$\micron$ image (see
fig.~\ref{fig:overscubw}), we do not use the 450~$\micron$ map.  The
need for deep sub-mm or mm observations of this region is highlighted
by the fact that both of the SCUBA maps have low S/N, suffer from
ambiguities in the spatial extent of the cloud due to the chopping
position, and have poorly understood spatial disagreements between the
850~$\micron$ and 450~$\micron$ observations.

\subsection{Optical Spectrum}

\begin{figure}
  \epsscale{0.95}\plotone{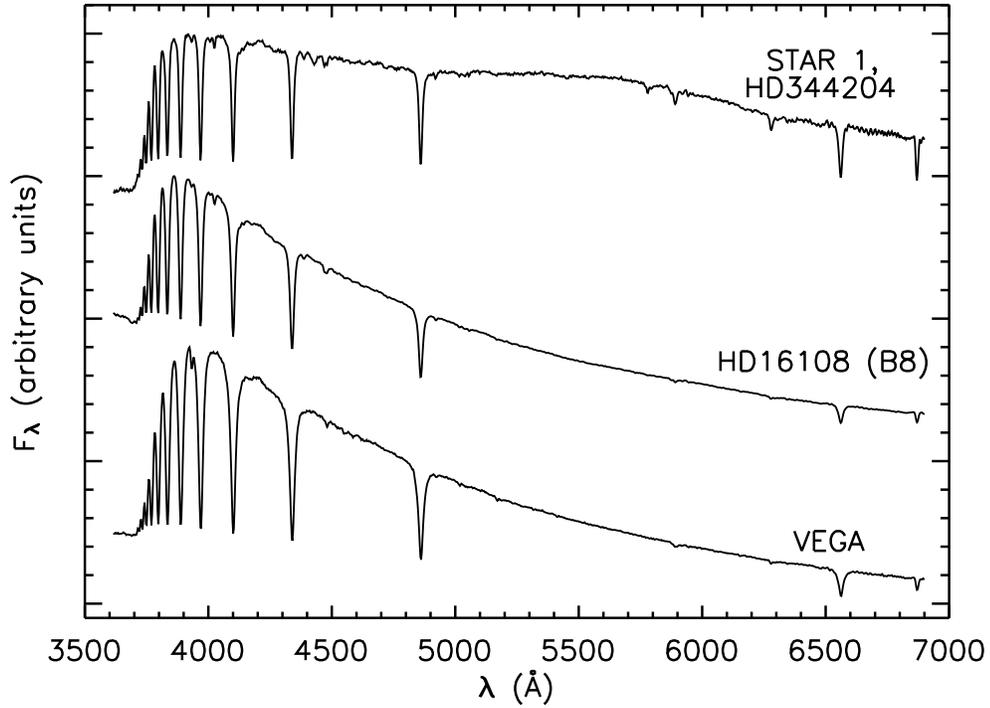}
  \caption{F$_\lambda$ vs. $\lambda$ for three stars: star 1
    (HD344204, top), HD1608 (middle), and Vega (bottom).  The data
    extend from 3615\AA\ to 6900\AA\ at 9\AA\ resolution.}
  \label{fig:sed}
\end{figure}

We have obtained optical spectra of three stars: HD344204 (star 1 in
figure~\ref{fig:overbw}), HD1608, and Vega.  These spectra cover the
optical range, from 3615\AA\ to 6900\AA\ with an effective resolution
of $\Delta \lambda$ = 9\AA, and were observed in July 2006 at the
Steward Observatory Bok 2.3m telescope at Kitt Peak using the Boller
and Chivens Spectrograph with a 400/mm grating in first order.  They
were processed with standard IRAF data reduction packages.  These data
are shown in figure~\ref{fig:sed} and discussed in \SS~3.1 and 3.2.

\section{Analysis}

\subsection{Two associated stellar sources}

\begin{figure}
  \epsscale{0.65}\plotone{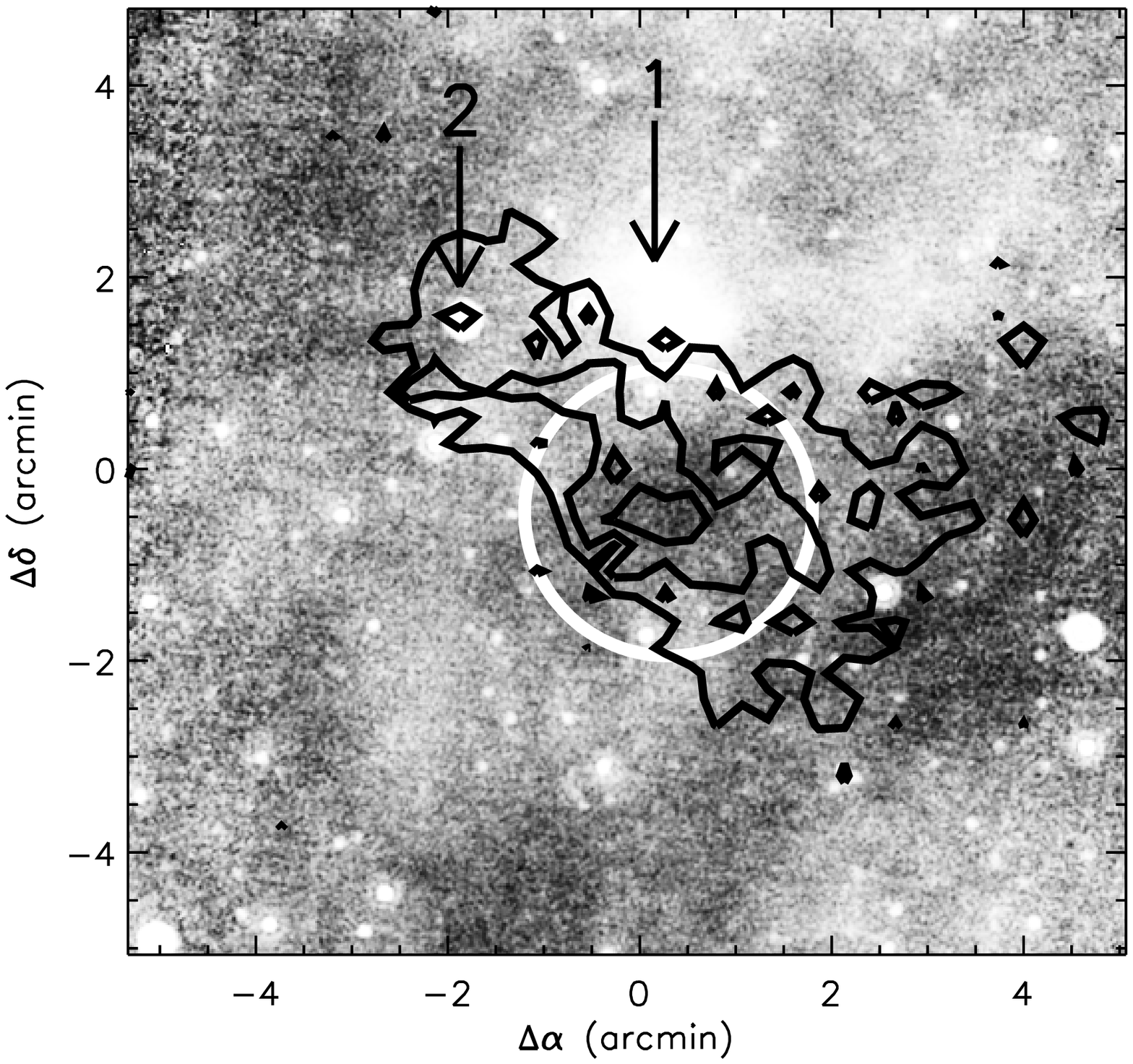}
  \caption{24~$\micron$ image with $^{13}$CO contours overlayed.  The
    image is $\sim\!10\arcmin$ on a side, is centered on the CO map at
    RA = $19^h 20^m 48^s$, Dec = $+23^o 29\arcmin 57\arcsec $, and is
    displayed at the original mosaic pixel scale of $1\farcs24$.  The
    $^{13}$CO contour levels are 0.8, 1.4, and 2.0 K.  The white
    circle is centered on the 24$\micron$ shadow and the peak of the
    $^{13}$CO emission.  The two arrows indicate two sources that are
    likely associated with \cbo; their broad-band SEDs are shown in
    fig.~\ref{fig:sed} and discussed in \S~3.1. }
  \label{fig:overbw}
\end{figure}

\begin{figure}
  \epsscale{0.95}\plotone{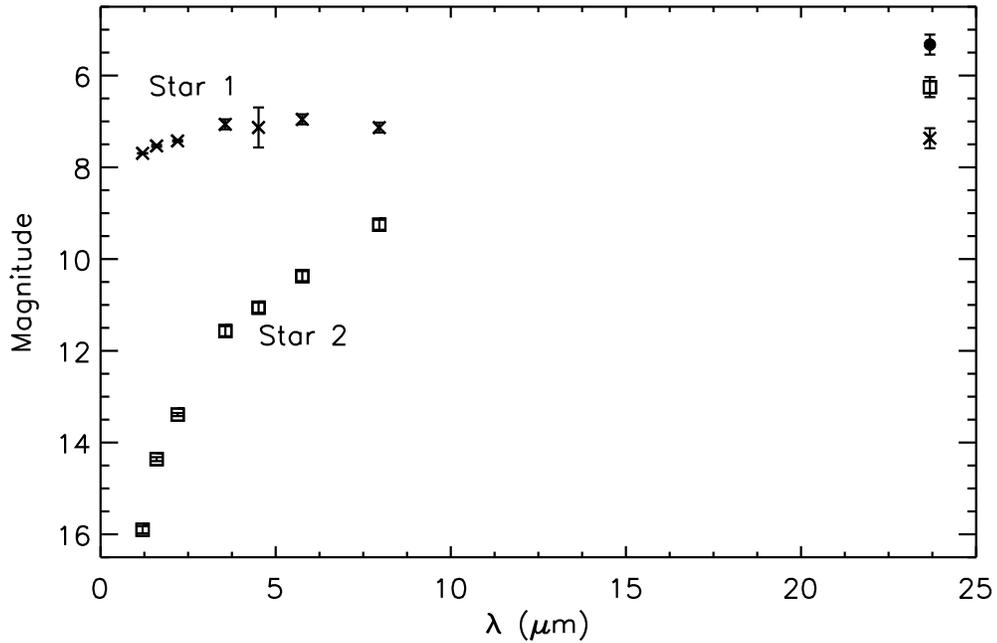}
  \caption{Broad-band SEDs of two sources associated with \cbo; we
    show the 2MASS, IRAC and MIPS magnitudes.  The $\times$'s show the
    magnitudes of star 1 (HD344204), at RA = $19^h 20^m 47^s$, Dec =
    $23^o 31\arcmin 40.637\arcsec$.  The solid circle indicates the
    magnitude of star 1 using a large aperture to include the dust
    emission surrounding the source.  The open squares show the
    magnitudes of star 2, at RA = $19^h 20m 57^s$, Dec = $23^o
    31\arcmin 37.6\arcsec$.  }
  \label{fig:phot}
\end{figure}

There are two bright sources in the 24$\micron$ image that are likely
associated with \cbo.  The first, labeled source 1 in
fig.~\ref{fig:overbw}, is a bright point source, just to the north of
the 24~$\micron$ shadow, with a large amount of diffuse emission.
This star is HD344204 (IRAS 19186+2325), and is located at RA = $19^h
20^m 47^s$, Dec = $23^o 31\arcmin 40.6\arcsec$.  The second star,
labeled source 2 in fig.~\ref{fig:overbw}, is spatially coincident
with a small peak in the $^{13}$CO emission, and is located at RA =
$19^h 20^m 57^s$, Dec = $23^o 31\arcmin 37.6\arcsec$.  The broad-band
SEDs of these two sources are plotted in fig.~\ref{fig:phot}, and
include 2MASS \citep{skrutskie06} J, H, and K data (see \S~2.3), IRAC
[3.6~$\micron$], [4.5~$\micron$], [5.8~$\micron$], and [8.0~$\micron$]
data, and MIPS 24~$\micron$ fluxes.  Although star 1 appears to be
indistinguishable from the surrounding diffuse emission in
figure~\ref{fig:overbw}, this is only due to the scale used to display
the image.  To measure the flux of the star while avoiding
contamination from the surrounding diffuse emission we use a very
small photometry aperture radius, $6.23\arcsec$, and sky annulus,
$6.23\arcsec$ to $7.47\arcsec$.  We derive an aperture correction of
2.1 to this flux using an isolated point source, measured with the
same aperture geometry as that listed above.  We compare this result
to the flux derived using a $13\arcsec$ aperture, a $20\arcsec$ to
$32\arcsec$ sky annulus, and the 24~$\micron$ aperture correction
recommended by the Spitzer Science Center.  For comparison
fig.~\ref{fig:phot} also indicates the 24~$\micron$ flux of Star 1
measured with a large aperture with radius $= 30\arcsec$ and an inner
and outer sky annulus radius $ = 34\arcsec$ and $38\arcsec$
respectively, meant to include all the light from the diffuse emission
surrounding the source.

In fig.~\ref{fig:phot} we also show the broad-band SED of source 2.
The IRAC colors are ([3.6~$\micron$]-[4.5~$\micron$]) = 0.5 mag, and
([5.8~$\micron$]-[8.0~$\micron$]) = 1.1 mag.  The models by
\citet{whitney03} suggest that these colors are consistent with those
of a late Class 0 source.

\subsection{The Distance}

To understand the nature of the \cbo\ 24~$\micron$ shadow one must
measure its physical properties, such as mass and size; to do so one
needs an accurate distance.  While star counts are commonly used to
estimate distances to nearby clouds, in the case of \cbo\ this method
is not reliable due to the small number of foreground sources.
Another distance estimator is the LSR velocity relation.  However,
this method is not reliable for nearby objects whose motions are still
locally dominated, as is the case of \cbo.  \citet{neckel80} estimate
the distance to this cloud to be $\sim\!400$~pc using the
discontinuity in A$_V$ with distance.  \citet{dame85} argue that this
distance is consistent with the narrow line width they measure and
therefore reject the other plausible distance to this cloud, that of
the Vul OB1 association at 2.3~kpc, noting that this longer distance
would imply a much larger mass and line width.

We can obtain a rough estimate of the distance using the colors of
source 2 (see fig.~\ref{fig:phot}) and the \citet{whitney03} models.
Their color-magnitude ([5.8~$\micron$]-[8.0~$\micron$]) vs.\
[3.6~$\micron$] relation for a face-on late Class 0 source, with
[3.6~$\micron] \sim 8$~mags, yields a distance $\sim\!700$~pc.  If we
assume a model with the same ([5.8~$\micron$]-[8.0~$\micron$]) color
but which is slightly more inclined, with a [3.6~$\micron] \sim
9.7$~mags, we obtain a distance $\sim\!330$~pc.  These distance
estimates are highly speculative, as the uncertainty in the magnitudes
makes the observed colors marginally consistent with later-type models.
Furthermore, these models are relatively untested.

In this context, i.e., determining a distance to \cbo, source 1 (see
fig.~\ref{fig:overbw}) draws attention for two reasons.  This star has
a large amount of diffuse emission at 24 and 70~$\micron$.  It is also
coincident with the truncation of the $^{13}$CO emission on the
northern edge of \cbo.  Based on these two facts we conclude that
source 1 is very likely to be associated with the cloud.  Our spectrum
of source 1 shows it to be a B7 star, based on the strong Balmer
absorption lines and HeI features that are evident in
fig.~\ref{fig:sed}.  Assuming this is a main-sequence star, its
distance is $\sim\!400$~pc, and if it is a giant (luminosity class
III) its distance is $\sim\!600$~pc.  The broad feature observed in
the spectrum of this source near 5700\AA\ may be due to the reddening
curve.  As a matter of historical interest, A. J. Cannon classified
this star as B9.  In the following analysis, where it is relevant, we
assume a distance of $400$~pc.

\subsection{Extinction law analysis}

\begin{figure}
  \epsscale{0.95}\plotone{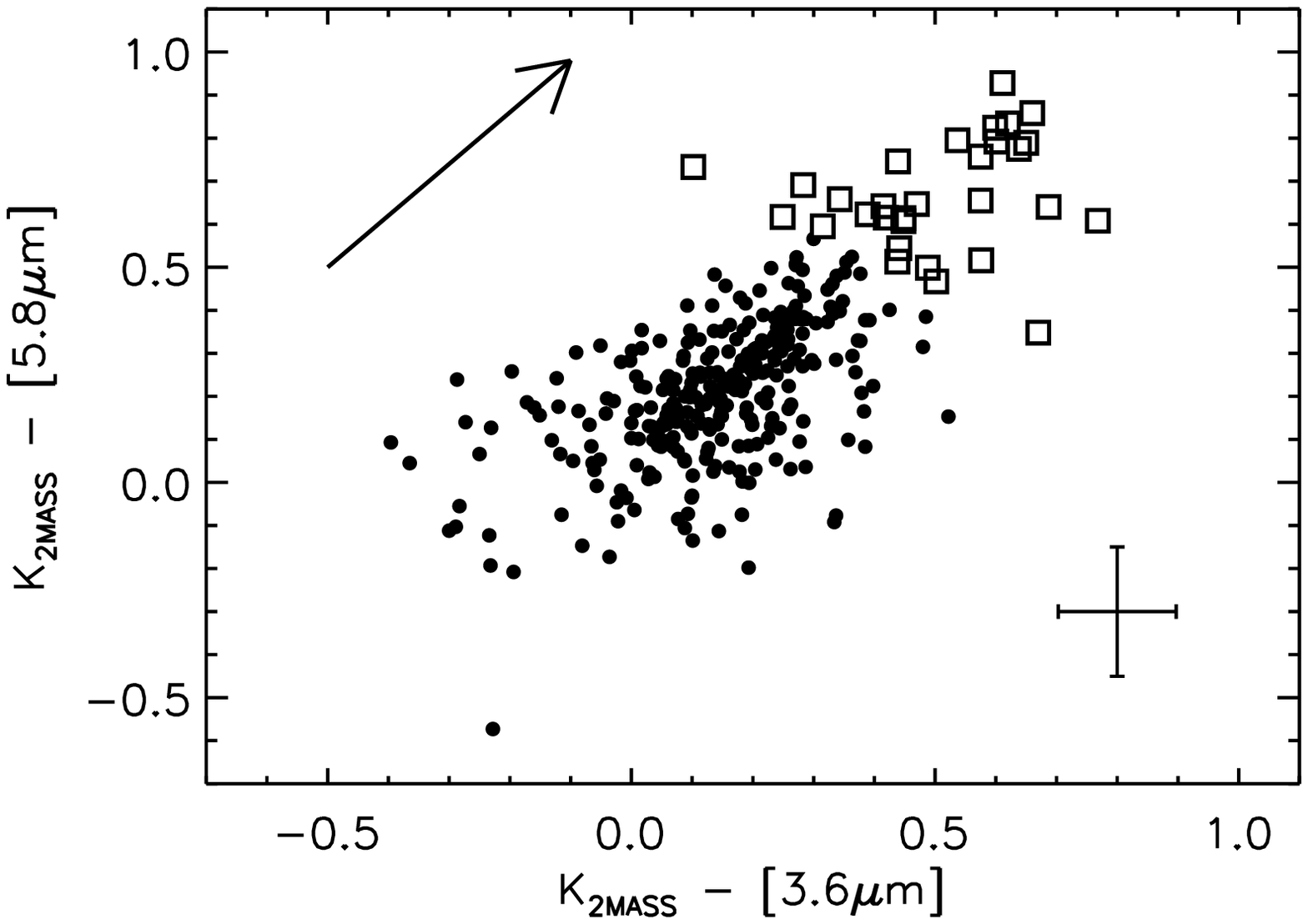}
    \caption{Color-color plot of stars observed towards \cbo.  We
      combine 2MASS K-band data with 3.6~$\micron$ and 5.8~$\micron$
      IRAC channels.  The squares indicate the reddened stars,
      selected with a 2-$\sigma$ clipping criterion, and the filled
      circles indicate the unreddened stars.  The black arrow shows
      the best-fit extinction vector slope of 1.2, in good agreement
      with the \citet{indebetouw05} extinction slopes.  The median
      errors for the data are shown in the lower right.  The
      extinction vector derived from the K-[3.6$\micron$]
      vs. K-[4.5$\micron$] colors also shows good agreement with
      \citet{indebetouw05}.  }
    \label{fig:irac2}
\end{figure}

We have used the IRAC and 2MASS data to probe the extinction law in
\cbo.  These data are used to generate the color-color plots shown in
fig.~\ref{fig:irac2}.  We calculate the errors in these colors by
adding the respective IRAC and 2MASS K-band errors in quadrature; the
median values of these errors are plotted in the lower right corner of
fig.~\ref{fig:irac2}.  To analyze the colors, we compare the best-fit
reddening vector to those measured by \citet{indebetouw05}.  We
consider stars to be reddened if their colors are more than $2 \sigma$
away from the mean colors of the ensemble.  These reddened stars are
indicated in fig.~\ref{fig:irac2} as open boxes.  We then fit the
slope of the reddening vector, including both x- and y-errors, using
the IDL routine {\it fitexy.pro} \citep{press92}.  We also include the
median value of the unreddened stars and assign it zero error.  The
resulting reddening vector slope is shown in fig.~\ref{fig:irac2} as a
solid black line.  Because this slope agrees reasonably well with the
\citet{indebetouw05} extinction vectors, we conclude that the dust
found in \cbo\ is not anomalous and that we are justified in using a
``standard'' reddening law to determine the dust properties in this
globule.  The agreement with the \citet{indebetouw05} extinction is
not surprising because, as was shown in \citet{harris78} and
\citet{rieke85}, there is generically almost no difference in the
infrared extinction in dense clouds even though the visual bands may
deviate significantly from their low-density values.  We note that we
exclude source 2, discussed in \S~3.1, from this analysis for two
reasons: first, because it has a very large 24~$\micron$ excess, and
second, because its broad-band SED is consistent with the colors of a
proto-star, being too red to be simply due to a foreground dust screen
and a normal star (see fig.~\ref{fig:phot}).

\subsection{The 24~$\micron$ shadow: Optical depth and column density profile}

\begin{figure}
  \epsscale{0.95}\plotone{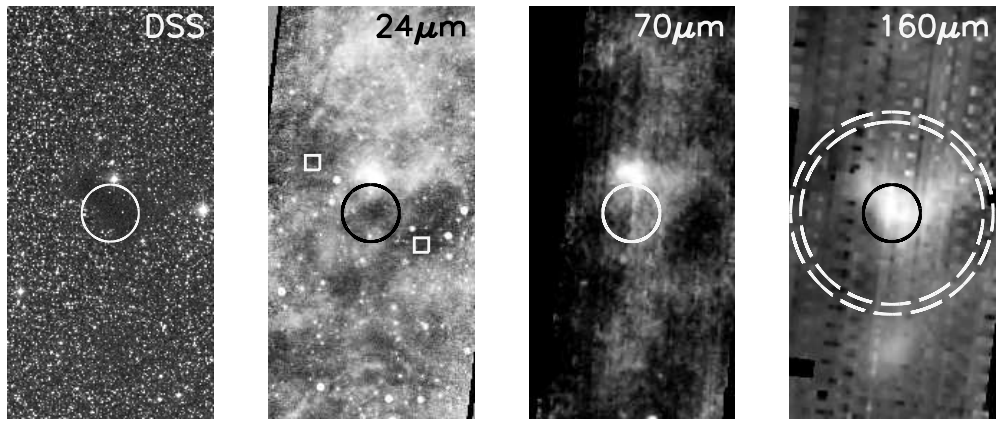}
  \caption{Gallery of \cbo\ data. The images are centered on the
    $24\micron$ shadow at RA = $19^h 20^m 48^s$, Dec = $+23^o
    29\arcmin 45\arcsec$, and are oriented such that north is up and
    east is to the left.  The images have a height of $24\arcmin$ and
    a width of $12\arcmin$.  The corresponding wavelengths are labeled
    in the top right corners.  The optical image is from the red
    Digital Sky Survey.  The original MIPS mosaic pixel scales are
    $\sim 1\farcs24$ pix$^{-1}$, $9\farcs8$ pix$^{-1}$, and
    $16\arcsec$ pix$^{-1}$, for 24~$\micron$, 70~$\micron$, and
    160~$\micron$ respectively.  The 24~$\micron$ image displayed here
    is binned down by a factor of $4$, and the DSS, $70$, and
    160~$\micron$ images are re-gridded to the same scale.  All four
    images are marked with a 100$\arcsec$ circle centered on the
    location of \cbo.  The 24~$\micron$ image is marked with two $\sim
    50\arcsec \times 50\arcsec$ boxes indicating the background
    regions used to estimate the overall uniform background level (see
    \S~3.3 for more details).  The 160~$\micron$ image shows the
    aperture used for the dust mass determination (see \S~3.2)
    indicated a black circle with a radius of $100\arcsec$, and the
    inner and outer sky annuli radii at $5\farcm3$ and $6\farcm4$,
    both marked as white dashed circles.}
  \label{fig:mips}
\end{figure}

The 24~$\micron$ image shows a shadow, or depression in the emission,
centered at RA = $19^h 20^m 48^s$, Dec = $+23^o 29\arcmin 45\arcsec$
(see fig.~\ref{fig:mips}), coincident with the location of the dark
cloud \cbo\ \citep[L771;\ e.g.,][]{clemens88,visser02}.  This shadow,
about 70$\arcsec$ in radius, is coincident with the peak in the
$^{12}$CO and $^{13}$CO maps (see figs.~\ref{fig:co} and
\ref{fig:overbw}), the 160~$\micron$ emission (see
fig.~\ref{fig:mips}), and the SCUBA emission (see
fig.~\ref{fig:overscubw}).  We do not observe a shadow at 70 and
160~$\micron$.  In fact, at 70~$\micron$ the cloud is at best only
marginally detected in emission and may be dominated by the light from
source 1 just to the north of the 24~$\micron$ shadow.  The shadow at
24~$\mu$m is a result of the cold and dense material in \cbo\ blocking
the background radiation.

In the following derivation of the density profile we correct for the
large-scale foreground emission at 24~$\micron$ (see $f_{DC}$ and
discussion below).  This large-scale component is likely to be
composed mainly of zodiacal light; if we underestimate it we will
effectively {\it wash out} the shadow signal and therefore
underestimate its optical depth profile, density profile, and mass.
Conversely, if we overestimate the large-scale background level, we
will also overestimate the optical depth profile.  Keeping this in
mind, we estimate the zodiacal contribution by using the darkest parts
of the image.  Therefore, strictly speaking, we are deriving an upper
limit to the optical depth profile.  This approach is conservative
because it allows us to derive a robust profile consistently and
independently, without having to use other data to set the
normalization of the density profile.  While it is possible that the
emission varies on shorter scales and, more specifically, emission
from \cbo\ fills in the 24~$\micron$ shadow, we consider this to be
unlikely due to the fact that the shadow is not detected at
70~$\micron$.  It is not plausible that \cbo\ would be emitting
significantly at 24~$\micron$ while remaining undetected at the longer
wavelength.  In the following discussion we describe the method used
to derive the optical depth and extinction profile for the
24~$\micron$ shadow.

First, we estimate the overall large-scale uniform background level in
the image, $f_{DC}$.  We do so using two dark regions in the image,
free from sources, indicated by boxes in fig.~\ref{fig:mips}, one to
the North-East and one to the South-West of the shadow, each one about
50$\arcsec$ on a side.  We use the first percentile flux value in
these boxes, $f_{DC} = -5.732$~mJy~arcsec$^{-2}$, as a lower limit on the
overall uniform level in the image.  We use this background level to
set the true image zero level by subtracting it from the original
image.  Then we mask out all bright sources in the image by clipping
all pixels with values $3 \sigma$ above the mean.  Finally, we use
this background-subtracted and bright-source masked image to derive an
optical-depth and extinction profile.  For completeness, we estimate
the error in $f_{DC}$ by simulating the pixel distribution in the two
dark regions.  We estimate $\sigma_{DC}$ by simulating $n_{pix} =
3362$ pixel values drawn from a normal density with mean and standard
deviation equal to those measured in the two regions and storing the
first percentile pixel value.  We repeat $10^4$ times, and calculate
the standard deviation in the simulated $f_{DC}$ of $\sigma_{DC} =
0.083$~mJy~arcsec$^{-2}$.

\begin{figure}
  \epsscale{0.95}\plotone{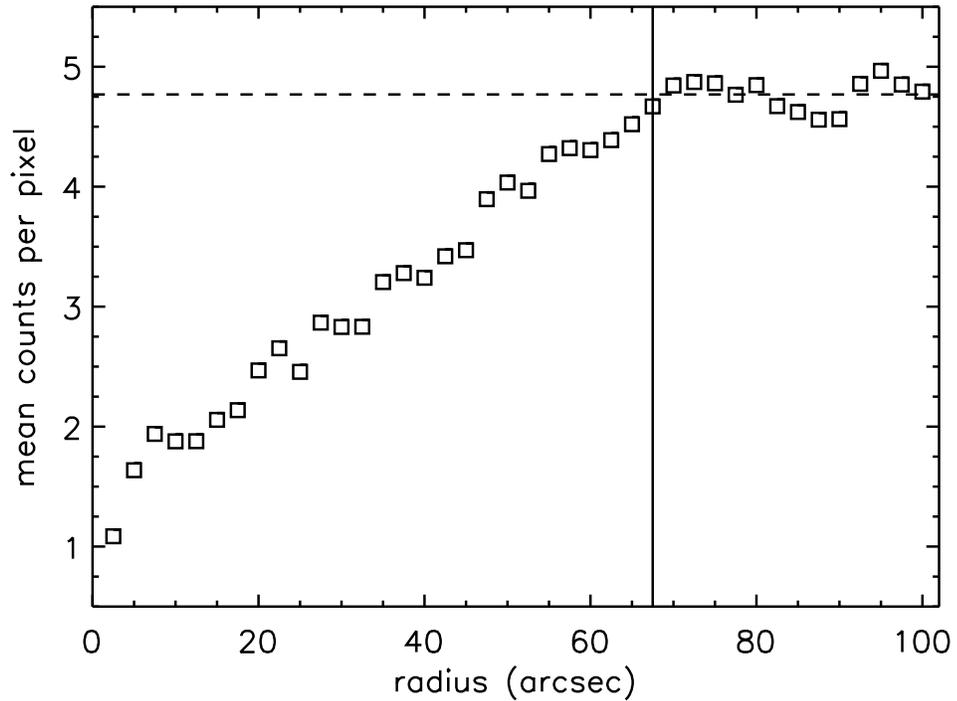}
  \caption{The radial flux profile for the $24$~$\micron$ shadow.  The
    squares indicate the mean counts per pixel in each $2\farcs5$
    annulus.  The chosen truncation radius for the shadow, at
    $67\farcs5$ is marked as a solid line; the local background flux
    level $I_0$ is marked with a dashed line.}
  \label{fig:cont}
\end{figure}

We proceed by dividing the shadow into regions of nested (concentric
and adjacent) annuli $2.5\arcsec$ in width, the inner-most region
being a circle with a radius of $2.5\arcsec$.  These regions are
centered on the darkest part of the shadow, which is also roughly
coincident with the peak in the $^{13}$CO emission.  We measure the
average flux per pixel in each region out to a radius of
$\sim\!100\arcsec$, chosen to be big enough to allow for a reasonable
estimate of the background immediately adjacent to the shadow.  We
show the derived radial profile in fig.~\ref{fig:cont}.  Based on this
profile, we set the boundary of the shadow to be at a radius of
$67\farcs5$ (marked in fig.~\ref{fig:cont} with a solid line) where
there is a flattening in the derived profile.  We estimate the
background flux by averaging the values of the annuli outside
$70\arcsec$ and within $100\arcsec$.  This background value is marked
in figure~\ref{fig:cont} with a dashed line.

We use this profile to calculate the optical depth $\tau_{24}$ of the
shadow.  In a given annulus $\tau_{24}$ is given by $\tau_{24} = -\ln
(I/I_0)$, where $I_0$ is the background level and $I$ is the shadow
flux.  The calculated value of $\tau_{24}$ varies from $\sim\!1.5$ at
the center to about 0.02 at a radius of $67\farcs5$.  The average mass
column density in each annulus is then given by
\begin{equation}
\Sigma = \frac{\tau_{24}}{\kappa_{abs,24}}f,
\end{equation}
where $f (= 100)$ is the gas-to-dust ratio, and $\kappa_{abs,24}$ is
the absorption cross-section per mass of dust.  In this work we use
the value of $\kappa_{abs,24} = 5.283 \times 10^{2}$~cm$^2$~gm$^{-1}$
calculated by \citet{draine03a,draine03b} from his $R_V = 5.5$ model.
The choice of a model with a high $R_V$ value relative to the diffuse
ISM is intended to account for some of the grain-growth effects likely
to be taking place in \cbo, as evidenced by the depletion indicated by
our $^{12}$CO and $^{13}$CO data.  This column density profile can be
converted to an extinction profile using the relation $N(H_2)/$A$_V =
1.87\times10^{21}$~atoms~cm$^{-2}$~mag$^{-1}$ \citep{bohlin78}, where
we assume that all of the hydrogen is in molecular form.  We use a
value of $R_V = 3.1$, appropriate for the diffuse ISM where this
relation was measured, to convert from $E(B - V)$ to $A_V$.  The
corresponding extinction is given by
\begin{equation}
A_{\rm V} = \frac{\Sigma}{1.87\times 10^{21}\,{\rm cm}^{-2}\,{\rm
    mag}^{-1} \mu_{\rm H_2} m_{\rm H}}, 
\end{equation}
where $\mu_{\rm H_2}$ is the effective molecular weight per hydrogen
molecule.  The molecular weight per hydrogen molecule is related to
the mass fraction of hydrogen, $\mathcal{M({\rm H})}/\mathcal{M}$
(where $\mathcal{M} = \mathcal{M({\rm H})} + \mathcal{M({\rm He})} +
\mathcal{M({\rm Z})}$), by
\begin{equation}
\mu_{\rm H_2} = \frac{\mathcal{M}}{m_{\rm H} N(\rm H_2)} =
\frac{2\mathcal{M}}{m_{\rm H} N(\rm H)} =
\frac{2\mathcal{M}}{\mathcal{M({\rm H})}}.
\end{equation}
For a cosmic hydrogen mass fraction $\mathcal{M({\rm H})}/\mathcal{M}
= 0.71$, $\mu_{\rm H_2} = 2.8$ (Jens Kauffmann - private
communication, 2006).  We consider two sources of uncertainty in the
column density profile for the 24~$\micron$ shadow: the DC-background
error and the uncertainty in our assumed dust model,
$\kappa_{abs,24}$.  We measure an error in the DC-background level of
$\sigma_{DC} = 0.083$~mJy~arcsec$^{-2}$ (described above).  We then
calculate the corresponding uncertainty in A$_V$ by propagating
$\sigma_{DC}$ through the equation for $\tau_{24}$ and scaling
appropriately.  We estimate the dust model uncertainty to be half of
the difference in $\kappa_{abs,24}$ between the \citet{weingartner01}
$R_V = 3.1$ model and the $R_V = 5.5$ model.  To obtain an estimate of
the total uncertainty in $A_V$, we add these two components and a
$10\%$ systematic error floor in quadrature.

\begin{figure}
  \epsscale{0.95}\plotone{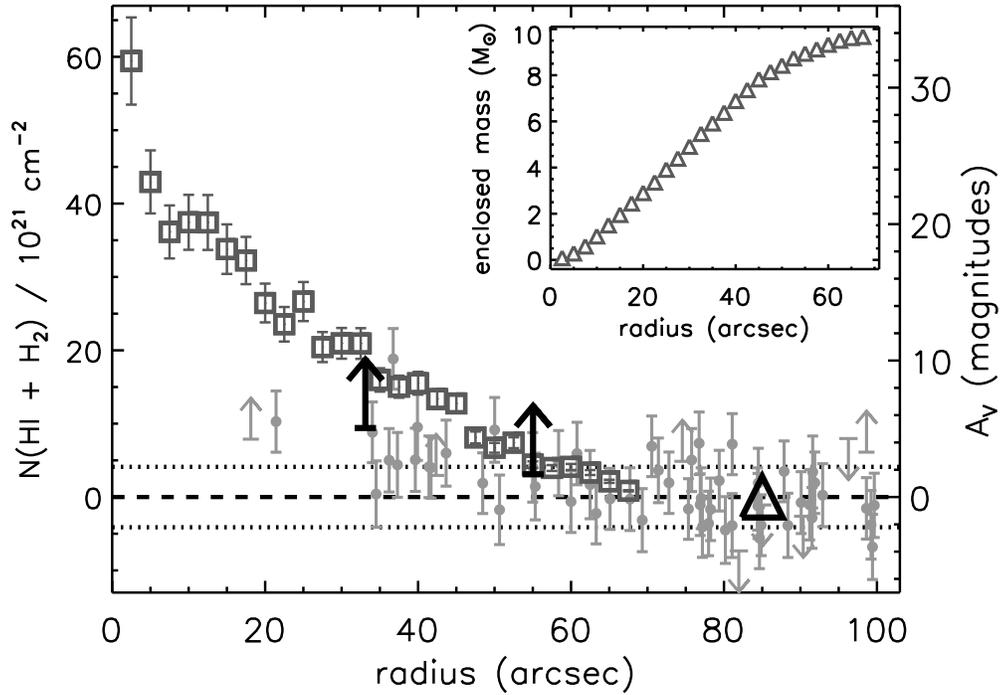}
  \caption{Comparison of the CB190 column density profiles: the
    $24$~$\micron$ shadow profile is shown with dark grey squares, and
    the individual $2$MASS stellar A$_V$ estimates are marked with
    light grey filled circles.  The two dotted lines indicate the
    1-$\sigma$ scatter in the 2MASS data just outside the globule,
    between 70$\arcsec$ and 100$\arcsec$.  The black symbols indicate
    the best-fit Gaussian mean values for the 2MASS data in three
    radial bins, inside 40$\arcsec$, 40$\arcsec$ to 70$\arcsec$, and
    70$\arcsec$ to 100$\arcsec$, and are plotted versus the average
    radius in each respective bin; the two inner-most points are shown
    as lower limits (black arrows) because of the inclusion of
    individual 2MASS lower limits in the best-fit Gaussian
    calculation.  The inset shows the corresponding enclosed mass for
    the $24$~$\micron$ extinction profile.  }
  \label{fig:av}
\end{figure}

The column density profile and corresponding errors are shown in
figure~\ref{fig:av} along with individual 2MASS point-source
extinction estimates.  We use 2MASS sources with data quality flags
better than, and including, ``UBB'', and include ``C'' quality
measurements in a bandpass when the other two filters are ``B''
quality or better.  Sources with upper limits in two or more bands are
rejected.  This selection ensures that we include the maximum number
of quality extinction measurements while not biasing the object
selection towards stars with lower amounts of reddening.  The
reddening of the accepted sources is measured using the $(J - K)$
color and the \citet{rieke85} extinction law: $A_V = E(J - K)/0.17$.
The ``intrinsic'' $(J - K)_0 = 1.34$ color of the stars is measured in
the 70$\arcsec$ to 100~$\arcsec$ annulus centered on the shadow, to
isolate the effects of the globule from that of nearby diffuse dust.
We reject sources between 0$\arcsec$ and 70$\arcsec$ with $E(J - K)$
values lower than $-0.37$, the 1-$\sigma$ value for the scatter in the
sources between 70$\arcsec$ and 100$\arcsec$. This 1-$\sigma$ range is
marked in figure~\ref{fig:av} with two dotted lines.  Two sources
fulfill this criterion, located at r $\sim\!25\arcsec, 67\arcsec$ from
the center.  We reject one other source with an A$_V \sim 3$ located
at $r \sim 9\arcsec$ on the basis that it is likely to be a foreground
object.  We calculate a probability of 1.4\% of finding three
foreground stars, using the stellar number densities from the Nearby
Stars Database \citep[NStars;][]{henry03}.  It is not surprising that
this probability is so low because the NStars catalog is not complete.
A more accurate estimate of the likelihood of finding three foreground
stars within the shadow is determined as follows.  In the 70$\arcsec$
to 100~$\arcsec$ annulus there are three stars with upper limits below
A$_V = 0$ and two stars with A$_V < -0.37$.  Furthermore, the
70$\arcsec$ shadow and the surrounding 70$\arcsec$ to 100~$\arcsec$
annulus have the same area.  We therefore expect $\sim\!4$ foreground
stars in the shadow region, and hence rejecting three stars is
reasonable.  Using the remaining stars we calculate the best-fit
Gaussian parameters for the distribution of $A_V$ values in two
annuli, from 0~$\arcsec$ to 40~$\arcsec$ and from 40~$\arcsec$ to
70$\arcsec$.  Lower limits are not treated differently than proper
detections in our fitting procedure.  Therefore, when we plot the mean
values from these fits versus average radius the values are shown as
lower limits (see fig.~\ref{fig:av}).  The trend of increasing column,
or A$_V$, with decreasing radius can be seen clearly.

In this section we have presented a new technique for analyzing
24~$\micron$ shadows which allows for a smooth estimate of the density
profile of these cold cloud cores at a 6$\arcsec$ resolution.  Most
importantly, this method traces gas and dust down to the densest
regions in the cloud.

\section{Mass estimates}

In the following mass estimates we assume a distance to \cbo\ of
400~pc (cf. \S~3.2).  We present a summary of these calculations in
Table~1.  

\subsection{24~$\micron$ shadow mass}

We use the optical depth profile derived in \S~3.4 to calculate the
mass of the 24~$\micron$ shadow.  The dust mass in a given annulus is
\begin{equation}
  M_d = \frac{\tau_{24}}{\kappa_{abs,24}} \Omega_{pix} n_{pix} D^2,
\end{equation}
where $\tau_{24}$ is the optical depth in the annulus,
$\kappa_{abs,24}$ is the absorption cross section per mass of dust at
$24\micron$, $\Omega_{pix}$ is the solid angle subtended by a pixel,
$n_{pix}$ is the number of pixels in the annulus and $D$ is the
distance to the cloud.  Using the $\kappa_{abs,24}$ from the
\citet{weingartner01} Milky Way synthetic extinction curve model with
$R_V = 5.5$, a gas-to-dust ratio $f=100$, and summing over all the
annuli within 70$\arcsec$, we derive a total 24~$\micron$ shadow mass
of $9.7\msun$ (see inset of fig.~\ref{fig:av}).  Using equation 4 and
standard propagation of errors we find that the fractional error in
the mass is equal to 3.3 times the fractional error in the local
background flux level, $I_0$ (marked as a dashed line in
fig.~\ref{fig:cont}).  For a fractional error in the background level
of $4\%$, corresponding to the the $95\%$ confidence interval for the
distribution plotted in fig.~\ref{fig:cont}, we calculate a $14\%$
error in the mass.  We note that this error analysis does not include
other significant systematic sources of error, like differences in
calculated dust opacities.  For example, \citet{ossenkopf94} calculate
a 24~$\micron$ $\kappa_{abs,24}$ = $8.69 \times
10^{2}$~cm$^2$~gm$^{-1}$ (for the thin ice mantle model generated at a
density of $n = 10^6$~cm$^{-3}$), a factor $\sim$ 1.6 greater than the
\citet{weingartner01} opacities used here, which would reduce our mass
to $\sim\!6\ \msun$.

We adopt a mass of $\sim\!10\ \msun$, although the estimates from the
160~$\micron$ data imply this value may be a lower limit.  This mass
is $\sim\!10$ times bigger than the mass derived from the CO
observations (see \S~5.1).  Furthermore, for a spherical cloud of the
same size and mass as \cbo, and a temperature of 10~K (consistent with
both the CO line data and the far infrared to submillimeter continuum
fit) we derive a Jeans mass $\sim\!4.1\ \msun$.  While such a low mass
would imply instability to collapse if we only consider thermal
pressure, this analysis neglects alternate forms of support -- namely
turbulence and magnetic fields -- which likely play a significant role
in \cbo.

\subsection{160~$\micron$ mass estimate}

Assuming that the dust is optically thin, the dust mass associated
with the 160~$\micron$ emission is given by
\begin{equation} 
  M_d = \frac{F_\nu D^2}{B_\nu(T) \kappa_{abs,160}},
\end{equation}
where D is the distance to the cloud and $\kappa_{abs,160}$ is the
absorption cross section per mass of dust at $160\micron$.  We use a
value for $\kappa_{abs,160} = 40.14$~cm$^2$~gm$^{-1}$, given by the
\citet{ossenkopf94} model with thin ice mantles and generated at a
density of $n = 10^6$~cm$^{-3}$.  These models account for grain
growth effects, which are likely to be taking place inside the cold,
dense cloud environment.  These authors show that the effects of grain
growth on the opacities are, however, minor.  To calculate the dust
mass we integrate the 160~$\micron$ flux within a 6.25~pixel
($100\arcsec$) radius aperture centered on the 24~$\micron$ shadow
(and the peak $^{13}$CO emission) and subtract an estimated pixel
background level.  We estimate this background per pixel by averaging
the flux in an annulus with an inner radius of 20~pixels and a width
of 4 pixels.  We assume a gas-to-dust ratio $f = 100$.  A typical
temperature for dust grains heated by the interstellar radiation field
(IRF) is $\sim\!18$~K \citep{draine03a}; if the cloud is externally
heated by the IRF then the total mass, given by equation 4, is
$\sim\!1.2\ \msun$.  However, we have shown in \S~2.5 that the
temperature of the cloud is in the range of 10.4K to 12K, giving a
total mass estimate between $\sim\!45\ \msun$ and $\sim\!14\ \msun$.
The \citet{ossenkopf94} models also include opacities for grains with
thick ice mantles; this opacity would scale these masses down by a
factor of 1.3.  We note that the \citet{weingartner01} $R_V = 5.5$ ISM
opacity of $\kappa_{abs,160} = 10.12$~cm$^2$~gm$^{-1}$ yields masses
that are larger by a factor of 4.  However, these models do not take
into account the grain growth effects that are thought to occur in
cold, dense regions.  We note that the temperature of CB190 is not
likely be constant throughout the cloud but instead probably decreases
inwards; this gradient would bias the 160$\micron$ observations toward
hotter dust on the outside of the globule and hence we might
underestimate the mass.

\subsection{CO mass estimate}

We use observations of $^{12}$CO and $^{13}$CO $J = 2 \rightarrow 1$
to estimate a mass for \cbo.  We use the $^{12}$CO data to estimate an
excitation temperature in the cloud; this temperature is then used in
conjunction with the $^{13}$CO data to estimate a mass, using standard
assumptions \citep[e.g.,][]{rohlfs04}.  We derive a $^{12}$CO gas
temperature $\sim\!10$~K.  For an isotope ratio of
[$^{12}$CO/$^{13}$CO] = 50 and a standard non-depleted ratio of
$n(CO)/n(H_2) = 1.5\times 10^{-4}$ \citep{kulesa05} we derive a cloud
mass $M = 4.2\ \msun$ by summing over the entire $\sim\!10'$ by $10'$
area of the image.  For an aperture with a radius of 70$''$ centered
on the peak of the $^{13}$CO emission (and the 24$\micron$ shadow) we
derive a total gas mass of $\sim\!1.1\ \msun$.  This CO mass is much
less than the $9.7\ \msun$ derived from the 24$\micron$ shadow profile
(cf.~\S~4.1).  These contradictory mass measurements can be reconciled
if freezeout is an important effect, which we consider to be a likely
scenario, or if the $n(CO)/n(H_2)$ ratio is over-estimated by about an
order of magnitude.  We note that masses derived in this fashion will
simply scale with the isotope ratio.  For comparison we calculate the
virial mass of the globule.  Using our higher resolution data we
measure the FWHM of the $^{13}$CO line to be $0.97$~Km~s$^{-1}$ at the
location of the shadow.  However, the C$^{18}$O data suggests that
there may be two velocity components broadening the $^{13}$CO line.
The line width of the C$^{18}$O observation is roughly $\Delta V
\sim\!0.5$~km~s$^{-1}$ at FWHM, in agreement with the NH$_3$
linewidth.  However, like the $^{13}$CO line, it appears to have some
asymmetry on the red side (see fig.~\ref{fig:hr12}).  If we estimate
the total width of the line by doubling the HWHM from the blue side,
then we obtain a $\Delta V \sim\!0.2$~km~s$^{-1}$.  We take
0.5~km~s$^{-1}$ as an upper limit on the line width and
0.2~km~s$^{-1}$ as a more plausible value.  These line widths
correspond to a virial mass between $18.0 \msun$ and $2.9 \msun$,
respectively.

\section{Bonner-Ebert models and possible support mechanisms}

When analyzing cold cloud density or extinction profiles it is common
to use theoretical density profiles to provide physical insight to the
system(s).  Bonnor-Ebert models are one such choice; they are
solutions to the isothermal equation, also known as the modified
Lane-Emden equation \citep{bonnor56,ebert55}.  The isothermal
equation, 
\begin{equation}
  \frac{d}{d\xi}\left(\xi^2\frac{d\Psi}{d\xi}\right) = \xi^2 e^{-\Psi},
\end{equation}
describes a self-gravitating isothermal sphere in hydrostatic
equilibrium.  Here, $\xi = r/r_c$ is the scale-free radial coordinate,
where
\begin{equation}
r_c = \left( \frac{kT}{4 \pi G \mu m_H\rho_c}\right)^{1/2}
\end{equation}
is the scale-radius, $r$ is the physical radial coordinate, $G$ is the
gravitational constant, $\mu$ (= 2.37) is the mean molecular weight
per free particle, $m_H$ is the mass of a hydrogen atom, $k$ is the
Boltzmann constant, $T$ is the temperature, and $\rho_c$ is the
central mass density. Finally,
\begin{equation}
  \Psi = -\ln(\rho/\rho_c),
\end{equation}
where $\rho$ is mass density as a function of radius.  Equation 7 can
be solved, with appropriate boundary conditions, to obtain the
scale-free radial profile of a Bonnor-Ebert cloud.  The singular
isothermal sphere ($ n \propto r^{-2}$) is a limiting solution, with
$\xi_{max} \rightarrow \infty$ \citep{chandra39}.

Any given solution to equation (7) is characterized by three
parameters: the temperature $T$, the central density $\rho_c$, and the
outer radius of the cloud $R$.  Once an outer radius is specified, a
model will be in an unstable equilibrium if $\xi_{max} > 6.5$, where
\begin{equation}
\xi_{max} =  R/r_c.
\end{equation}
Equivalently, an unstable model will have a density contrast between
the center and the edge of the cloud greater than 14.3
\citep{ballesteros03}.  For more discussion of Bonnor-Ebert models
see, e.g., \citet{evans01}, \citet{ballesteros03}, \citet{harvey03},
\citet{lada04} and \citet{shirley05}.

\subsection{Bonnor-Ebert fits}

We generate a set of Bonnor-Ebert models over a 3-D grid in
temperature, outer radius, and central density.  Our temperature grid
ranges from T = $8.0$~K to $98.0$~K in steps $\Delta T = 2.0$~K.  Our
radial grid varies from $\theta_{max} = 45\farcs0$ to $95\farcs0$ in
steps of $\Delta\theta_{max} = 2\farcs0$, and we assume a distance of
$400$~pc.  For the central density grid, we vary the $\xi_{max}$
parameter from $\xi_{max} = 3.0$ to $19.0$, in steps of $\Delta
\xi_{max} = 0.2$.  As can be seen from equations (8) and (10), at fixed
temperature and $\theta_{max}$ ($= R/D$, where D = 400~pc is fixed),
varying $\xi_{max}$ is equivalent to varying the central density
$\rho_c$.  We then integrate our calculated density profiles to obtain
column density profiles:
\begin{equation}
N_{BE}(r) = 2 \times \int^R_r \rho(r')\frac{dr'}{\sqrt{1 - (r/r')^2}},
\end{equation}
where $r$ is the projected distance from the center of the shadow and
the integration is along the line of sight $r'$.

We calculate the best-fit model by finding the minimum $\chi^2$ over
the grid in temperature, $\theta_{max}$, and $\xi_{max}$, where
\begin{equation}
\chi^2 = \sum_i \frac{(N_{i} - N_{BE,i})^2}{\sigma_{i}^2}.  
\end{equation}
Here, $N_{i}$ is the measured column density, $\sigma_{i}$ is the
corresponding uncertainty in $N_{i}$, $N_{BE,i}$ is the Bonnor-Ebert
model extinction, and the sum over $i$ represents the integration over
the spatial coordinate.  Errors in the best-fit parameters are
calculated using a Monte Carlo approach; we generate a set of $10^{3}$
A$_V$ extinction profiles, given by equation 3, with values of
$f_{DC}$ and $\kappa_{abs,24}$ with normal distributions, described in
\S~3.4.  We also include the $10\%$ systematic noise floor in our
calculation of the mock A$_V$ profiles.  For each one of these
artificial column density profiles we fit a model according to the
procedure outlined above.  The errors in the fitted parameters that we
quote are the standard deviations in the resulting best-fit parameter
distributions.

\begin{figure}
  \epsscale{0.95}\plotone{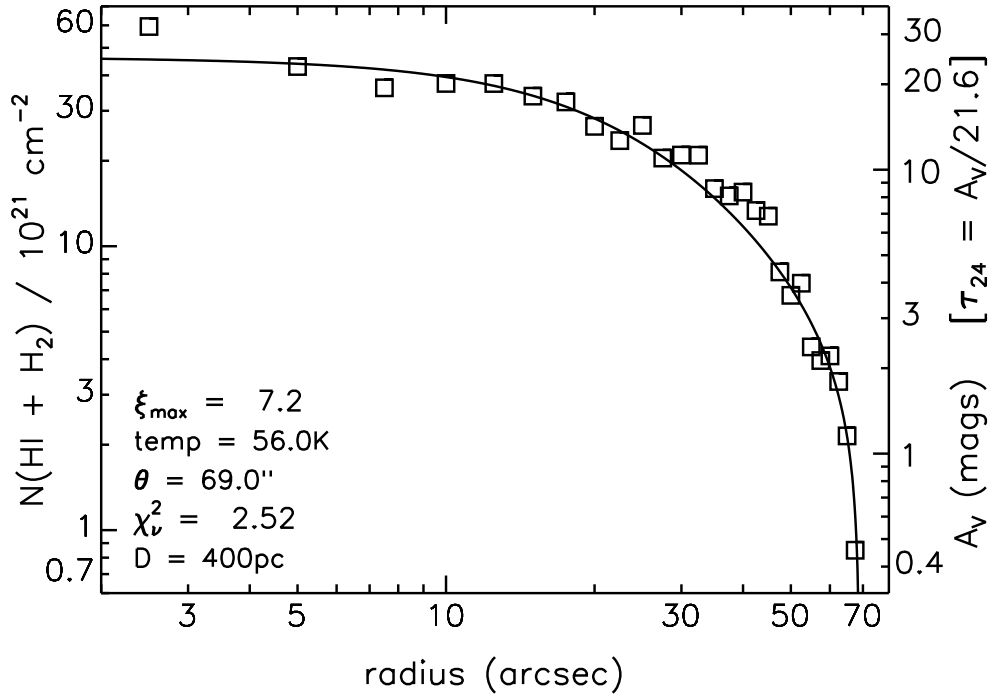}
  \caption{The $24$~$\micron$ column density profile (squares) is
  shown with the best-fit Bonnor-Ebert model (solid line).  The errors
  in the data are about the size of the squares.  The best-fit model
  parameters are indicated: temperature T $= 56.0$~K, $\theta_{max} =
  69\farcs0$, and $\xi_{max} = 7.2$.  The right-hand-side vertical
  axis indicates the magnitudes of visual extinction, or equivalently,
  $\tau_{24}$, the optical depth at 24~$\mu$m.}
  \label{fig:fit}
\end{figure}

We find a best-fit Bonnor-Ebert model with a temperature $T = 56.0 \pm
7.3$~K, $\theta_{max} = 69\farcs0 \pm 0.1$, and $\xi_{max} = 7.2 \pm
0.2$.  The central density of this model is $n_c = 1.78 \times
10^5$~cm$^{-3}$.  We show this model with the data in
fig.~\ref{fig:fit}.  Despite the reasonably good agreement between the
model and the 24~$\mu$m profile, this fit is inconsistent with the
molecular data and the SCUBA data: the temperature from the CO line is
$\sim\!10$~K while the continuum fit yields a temperature of 10 to
12~K.  We note that reducing the assumed distance to \cbo\ lowers the
best-fit model temperature: a distance of 200~pc brings the model
temperature down to $\sim\!28$~K.  However, we rule out this short
distance based on the likely association of star 1 with \cbo, see
fig.~\ref{fig:overbw}.  We discuss this distance in more detail in
\S~3.2.  Even though the best-fit $\xi_{max}$ is only slightly greater
than 6.5, our best-fit models always have either inconsistently high
temperatures or unacceptably small distances.  Therefore we must rule
out stable Bonnor-Ebert profiles as accurate representations of \cbo.

\subsection{Turbulence}

The temperature of \cbo\ is $\sim\!$10~K, based on our 160~$\micron$,
850~$\micron$, and CO observations, typical for cold cores
\citep{lemme96,hotzel02}.  This temperature corresponds to a thermal
line width of $\Delta V_{th} \sim 0.09$~km~s$^{-1}$ for C$^{18}$O.  As
discussed in \S~4.3, the observed line width of the C$^{18}$O
observation is between 0.2~km~s$^{-1}$ and 0.5~km~s$^{-1}$, much
broader than that expected from thermal support.  We take
0.5~km~s$^{-1}$ as an upper limit on the turbulent line width and
0.2~km~s$^{-1}$ as a more plausible value.

From \citet{hotzel02}, the energy contributed by turbulence to the
support of the cloud is
\begin{equation}
  E_{turb} = \frac{\mu_p}{30}\left( \frac{\Delta V^2}{\Delta V_{th}^2}
  - 1 \right)E_{th}
\end{equation}
Assuming that $\Delta V = \{0.5,0.2\}$~km~s$^{-1}$ then 
$E_{turb}/E_{th} = \{3.5,0.5\}$.  In this case, the maximum amount of
energy is $E_{tot} = 4.5E_{t}$, which would supply support equivalent
to a temperature of $45$K while the more plausible value is only
$\sim\!10-15$~K.  These temperatures are lower than our Bonnor-Ebert
best-fit value of $T = 56.0 \pm 7.3$~K.  Although turbulence might in
fact be a significant source of support in \cbo, it is not likely to
provide enough outward pressure to prevent collapse.

\subsection{Magnetic support}

Here we consider the effects of magnetic fields, an alternative to
turbulence as a support mechanism in \cbo.  Under a broad range of
conditions, the magnetic pressure is $B^2/8\pi$ \citep{boss97}. From
\citet{stahler05}, the mass that can be supported given a cloud radius
$R$ and magnetic field magnitude $B$ is
\begin{equation}
M = 70\msun \left( \frac{B}{10\mu G} \right) \left( \frac{R}{1pc}
\right)^2.
\end{equation}
\cbo\ has an average column density of $\langle N \rangle = 2 \times
10^{22}$~cm$^{-2}$, which corresponds to a line-of-sight magnetic
field of $B_{los} = 50$~$\mu G$, given by the observed correlation for
17 clouds with confirmed magnetic field detections \citep{basu04}.  We
note that this relation does have a large scatter of about 0.2 dex.
Assuming equipartition, the total magnitude of the expected magnetic
field is then $B_{tot} = \sqrt 3 \times B_{los} = 87$~$\mu G$.  Using
a radius of $R = 0.13$~pc, equation 13 gives a mass $M = 11$~$\msun$.
Therefore, the magnetic field could be of sufficient magnitude that it
may support this cloud and retard collapse.  This result is
interesting because magnetic support is often overlooked in studies of
cold cloud cores, yet in the case of \cbo\ it may play a dominant
role.  We note that a decrease of 0.2~dex in the magnetic field will
make it insufficient to support \cbo.

\subsection{Comparison with other globules}

\citet{kandori05} summarize observations of the density structure of
dark globules. They show that, of 11 starless cores with good density
measurements, 7 have profiles consistent with purely thermal support.
\citet{teixeira05} report three additional starless cores in Lupus 3,
of which two appear to be supported thermally. That is, of 14 such
cores, 9 have profiles consistent with pure thermal support. In
general, the five cores where an additional support mechanism is
required do not have adequate line width measurements to assess the
role of turbulence.

Our study of \cbo\ places it among the relatively rare class of cores
that cannot be supported purely thermally. Our high resolution line
measurements indicate that the turbulence in this core is inadequate
for support also. It is plausible that the magnetic field in the
globule supplies the deficiency, if it is at an average level measured
for other cold cloud cores. If \cbo\ is currently supported in this
way, it is at an interesting phase in its evolution.  For the
properties of this cloud, the ambipolar diffusion timescale is
$\sim\!3 \times 10^6$ yr \citep{stahler05}, which is about a factor of
10 longer than the free-fall collapse timescale.  Thus, it is
predicted that magnetically supported cores are unstable over about
ten million years, as their magnetic fields leak out through ambipolar
diffusion
\citep[e.g.,][]{crutcher94,boss97,indebetouw00,sigalotti00,tassis04}.
At that point, we would expect \cbo\ to begin collapsing into a
star. If the magnetic field is lower than average, this process may
already have begun.

\section{Summary}

We have combined Spitzer MIPS and IRAC data with HHT and GBT
millimeter data of \cbo\ and arrive at the following conclusions:

\noindent $\bullet$ We introduce a method for studying the structure
of cold cloud cores from the extinction shadows they cast at
24~$\micron$.

\noindent $\bullet$ We derive an $A_V$ profile of the 24~$\micron$
shadow that is in good agreement with the reddening estimates derived
from the 2MASS data at the outer edges and reaches a maximum value of
$\sim\!32$ visual magnitudes through the center.  

\noindent $\bullet$ The mass measured from the optical depth profile
is a factor of $\sim\!2$ greater than the Jeans mass for this object.

\noindent $\bullet$ We fit Bonnor-Ebert spheres to our A$_V$ profile
and find that the best-fit temperatures are in contradiction with the
CO observations and the thermal continuum data, which indicate much
lower temperatures for this globule.  We also show that turbulence is
probably inadequate to support the cloud.  However, magnetic support
may be enough to prevent collapse.

\noindent These pieces of evidence together form a consistent picture
in which \cbo\ is a cold dark starless core.  Although collapse
cannot be halted with thermal and turbulent support alone, the
magnetic field may contribute enough energy that it could support
\cbo\ against collapse.  Hence, magnetic field support should be
included in evaluating the stability of other cold cloud cores.  \cbo\
appears to be at an interesting evolutionary phase.  It may be in the
first stages of collapse (if the magnetic field is weaker than
average).  Alternately, if it is currently supported by magnetic
pressure, it is expected that collapse may begin in some ten million
years as the magnetic field leaks out of the globule by ambipolar
diffusion.

\acknowledgements
The authors thank Charles Lawrence for use of the IRAC GTO time for
this project.  We also thank Michael M. Dunham for providing
unpublished data and Claire J. Chandler for sharing the reduced SCUBA
data.  We also thank Neal J. Evans II for useful discussion and
helpful comments.  AMS thanks Martin E. Pessah for helpful
discussions.  This publication makes use of data products from the Two
Micron All Sky Survey, which is a joint project of the University of
Massachusetts and the Infrared Processing and Analysis
Center/California Institute of Technology, funded by the National
Aeronautics and Space Administration and the National Science
Foundation.  Portions of this work were carried out at the Jet
Propulsion Laboratory, California Institute of Technology, under
contract with the National Aeronautics ans Space Administration.  This
work was supported by contract 1255094 issued by Caltech/JPL to the
University of Arizona.


\begin{deluxetable}{lccr}
\tabletypesize{\scriptsize}
\tablecaption{Summary of mass estimates}
\tablewidth{0pt}
\tablehead{
\colhead{Data} 
& \colhead{Temp. [K]}
& \colhead{Aperture size [$\arcsec$]}
& \colhead{Mass [$\msun$]} 
}
\startdata
24$\micron$ & \nodata &  \phn70 & 10\ \ \ \ \ \ \\
Jeans Mass & 10 &  \phn70 &  4\ \ \ \ \ \ \\
160$\micron$$^a$ & 10 & 100 & $\sim$ 45\ \ \ \ \ \ \\
160$\micron$$^a$ & 12 & 100 &  $\sim$ 14\ \ \ \ \ \ \\
$^{12}$CO \& $^{13}$CO& \nodata & \phn70 &  $\geqslant$ 1$^*$\ \ \ \ \\
C$^{18}$O virial mass$^b$ & \nodata & \phn70 &  3\ \ \ \ \ \ \\
\enddata
\label{tab1}
\tablenotetext{*}{Evidence for CO freezout implies that this value is
  a lower limit.}
\tablenotetext{a}{Indicated as upper limits to allow for grain growth
  in dense regions.}
\tablenotetext{b}{Based on an estimated line width of 0.2~km~s$^{-1}$.}
\end{deluxetable}


\end{document}